\documentstyle[preprint,revtex,eqsecnum]{aps}
\begin{document}
\newtheorem{theorem}{Theorem}
\newtheorem{lemma}{Lemma}
\def\qed{\hfill \vrule height 7pt width 7pt depth 0pt \smallskip}
\def\up#1{\leavevmode \raise .3ex\hbox{$#1$}}
\def\down#1{\leavevmode \lower .5ex\hbox{$\scriptstyle#1$}}
\def\chiJ{\up{\chi}_{\down{J}}}
\hfill{ITD 92/93--9}\par
\draft
\begin{title}
Level-Spacing Distributions and the Airy Kernel
\end{title}
\author{Craig A.~Tracy\footnotemark[1]}
\footnotetext[1]{e-mail address: catracy@ucdavis.edu}
\begin{instit}
Department of Mathematics and Institute of Theoretical Dynamics,\\
University of California,
Davis, CA 95616, USA
\end{instit}
\author{Harold Widom\footnotemark[2]}
\footnotetext[2]{e-mail address: widom@cats.ucsc.edu}
\begin{instit}
Department of Mathematics,\\
University of California,
Santa Cruz, CA 95064, USA
\end{instit}
\begin{abstract}
Scaling level-spacing distribution functions in the ``bulk of the
spectrum'' in random matrix models of $N\times N$ hermitian matrices
and then going to the limit $N\rightarrow\infty$, leads to the
Fredholm determinant of the {\it sine kernel\/} $\sin\pi(x-y)/\pi (x-y)$.
Similarly a  scaling limit at the ``edge of the spectrum''
leads to the {\it Airy kernel\/} $\left[ {\rm Ai}(x) {\rm Ai}^\prime(y)
-{\rm Ai}^\prime(x) {\rm Ai}(y)\right]/(x-y)$. In this paper
we derive analogues
for this Airy kernel of the following properties of
the sine kernel: the completely integrable system of P.D.E.'s
found by Jimbo, Miwa, M{\^o}ri and Sato; the expression, in the
case of a single interval, of the Fredholm determinant in terms
of a Painlev{\'e} transcendent; the existence of a commuting
differential operator; and the fact that this operator can be
used in the derivation of asymptotics, for general $n$, of the
probability that an interval contains precisely $n$ eigenvalues.
\end{abstract}
\vfill\eject
\section{INTRODUCTION AND SUMMARY OF RESULTS}
\label{sec:intro_summary}
\subsection{ Introduction}
\label{subsec:intro}
In this paper we present  new results
for the {\it level spacing distribution functions\/}
obtained from scaling
 random matrix models of $N\times N$ hermitian matrices
at the edge of the support of the (tree-level) eigenvalue
densities when the parameters of the potential $V$
are  not ``finely tuned.''  This universality class is already
present in the Gaussian Unitary Ensemble.
It is known \cite{bowick_brezin,forrester,mehta_book,moore} that these
distribution functions are expressible in terms of a Fredholm
determinant of an integral operator $K$ whose kernel involves
 Airy functions.
\par
  It turns out that there  are striking analogies between the
properties of this   {\it Airy kernel\/}
\[ K(x,y) = {A(x) A^\prime(y) - A^\prime(x) A(y) \over x-y}
\]
where
$ A(x) = \sqrt{\lambda}\> {\rm Ai}(x)$,
and the {\it sine kernel\/}
\[ {\lambda\over \pi}\> {\sin\pi(x-y)\over x-y} \, ,  \]
whose associated Fredholm determinant describes the
classical level spacing distribution  functions first studied
by Wigner, Dyson, Mehta, and others \cite{dyson62a,mehta_book,porter}.
\par
To be a little more explicit about how these kernels arise we remind
the reader that in the GUE the probability density that the eigenvalues
of a random $N\times N$ hermitian matrix lie in infinitesimal
intervals about $x_1,\cdots, x_N$ is given by \cite{mehta_book}
\[ P_N(x_1,\cdots,x_N)={1\over N!} {\rm det}\left( K_N(x_i,x_j)\right)_{
i,j=1,\cdots,N} \]
where
\[ K_N(x,y)=\sum_{k=0}^{N-1} \varphi_k(x) \varphi_k(y) \, . \]
Here $\lbrace \varphi_k(x) \rbrace$ is the sequence obtained
by orthonormalizing the sequence
 \[\left\lbrace x^k \exp (-x^2) \right\rbrace\]
over $(-\infty,\infty)$.
  More generally
the $n$-point correlation function $R_n(x_1,\cdots,x_n)$, which is
the probability density that $n$ of the eigenvalues
(irrespective of order)  lie in
infinitesimal intervals about $x_1,\cdots,x_n$, is given by
\[ R_n(x_1,\cdots,x_n) ={\rm det}\left( K_N(x_i,x_j)\right)_{i,j=1,\cdots,n}\]
In particular $R_1(x)$, the density of eigenvalues at $x$, equals
$K_N(x,x)$.
\par
Replacing $K_N(x,y)$ by
\[ {1\over R_1(z)} K_N\left(z+{x\over R_1(z)},z+{y\over R_1(z)}\right)\]
has the effect of making the point $z$ in the spectrum the new origin
and rescaling so that the eigenvalue density at this point
becomes equal to $1$.  Now if $z$ is fixed then
\[ R_1(z)=K_N(z,z) \sim {1\over\pi}\sqrt{2N} \]
as $N\rightarrow\infty$, and
\[ \lim_{N\rightarrow\infty} {\pi\over \sqrt{2N}}
K_N\left(z+{\pi x\over \sqrt{2N}},z+{\pi y\over \sqrt{2N}}\right)=
{1\over\pi}\, {\sin\pi (x-y)\over  (x-y)}\, , \]
so we obtain the sine kernel in this way.\ \ (This limit is
obtained by applying the Christoffel-Darboux formula to $K_N(x,y)$
and then using the known asymptotics of $\varphi_N(x)$ as
$N\rightarrow\infty$.)\ \ The ``edge of the spectrum''
 corresponds to $z\sim\pm\sqrt{2N}$
and one has there the scaling limit \cite{forrester,moore}
\[ \lim_{N\rightarrow\infty} {1\over 2^{1/2} N^{1/6}}
K_N\left(\sqrt{2N}+{x\over 2^{1/2} N^{1/6}},\sqrt{2N}+{y\over 2^{1/2} N^{1/6}}
\right)= {{\rm Ai}(x){\rm Ai}^\prime(y) - {\rm Ai}^\prime(x) {\rm Ai}(y)\over
x-y}\, . \]
\par
We now present  some  of the
analogies  which we have found.  (That such analogies between properties
of the sine kernel and Airy kernel exist is perhaps
not suprising since they are both scaling limits of the same family
of kernels.)\ \
The first is the analogue  of the completely integrable system of
P.D.E.'s of Jimbo, Miwa, M{\^o}ri, and Sato~\cite{jmms}
 when  the underlying
domain is a union of intervals.
  The second is the fact that in
the case of the semi-infinite interval $(s,\infty)$ (the analogue
of a single finite interval for the sine kernel)
 the Fredholm determinant
is closely related to a Painlev{\'e} transcendent of the second
kind (the fifth transcendent arises for the sine kernel~\cite{jmms}).
  And the
third is the existence of a  second order differential
operator commuting  with the Airy operator $K$.  (The existence of such
a differential operator in the sine kernel case  has been known
for some time, see e.g.\ \cite{ince}, page 201.)\
 This last fact  leads to
an explicit asymptotic formula, as
the interval $(s,\infty)$ expands,  for the probability that it
contains precisely $n$ eigenvalues ($n=1,2,\ldots$)  (the analogue  of
results in \cite{btw,widom92}).  Here are our results, stated in some
detail.  (A short announcement  appears in \cite{tw2}.)
\subsection{  The System of Partial Differential Equations}
\label{subsec:sum_pde}
We set
\[ J=\bigcup_{j=1}^m\left(a_{2j-1},a_{2j}\right)  \]
and write $D(J;\lambda)$ for the Fredholm determinant of $K$ acting
on $J$.  We think of this as a function of $a=(a_1,\ldots,a_{2m})$.
Then
\begin{equation}
 d_a \log D(J;\lambda) = - \sum_{j=1}^{2m} (-1)^j R(a_j,a_j)\, da_j
\label{dlog_tau}
\end{equation}
where $R(x,y)$ is the kernel of the operator $K(I-K)^{-1}$ and
$d_a$ denotes exterior differentiation with respect to the $a_j$'s.
We introduce quantities
\begin{equation}
 q_j=(I-K)^{-1} A(a_j)\, , \ \ p_j=(I-K)^{-1} A^\prime(a_j)\, ,
\label{qp_defn}
\end{equation}
(which are the analogue  of the
quantities $r_{\pm j}$ of \cite{jmms};  see also
\cite{dyson92,its90,mehta92b,mehta_mahoux})
as well as two further quantities
\begin{equation}
 u=\left(A,(I-K)^{-1}A\right)\, , \ \ v=\left(A,(I-K)^{-1}A^\prime\right)
\label{uv_defn}
\end{equation}
where the inner products refer to the domain $J$.  Then the equations read
\begin{eqnarray}
{\partial q_j\over\partial a_k}&=& (-1)^k\,  {q_j p_k - p_j q_k\over
a_j-a_k}\, q_k\,\ \ \ (j\neq k), \label{airy_1}  \\
{\partial p_j\over\partial a_k}&=& (-1)^k\,  {q_j p_k - p_j q_k\over
a_j-a_k}\, p_k \ \ \  (j\neq k), \label{airy_2} \\
{\partial q_j\over\partial a_j}&=&
-\sum_{k\neq j} (-1)^k\,  {q_j p_k - p_j q_k \over a_j - a_k}\, q_k
+p_j - q_j u\, ,\label{airy_3} \\
{\partial p_j\over\partial a_j}&=&
-\sum_{k\neq j} (-1)^k\,  {q_j p_k - p_j q_k \over a_j - a_k}\, p_k
+a_j q_j + p_j u-2 q_j v\, , \label{airy_4} \\
{\partial u\over\partial  a_j}&=&(-1)^j q_j^2\, , \label{airy_5}  \\
{\partial v\over\partial  a_j}&=& (-1)^j p_j q_j \, . \label{airy_6}
\end{eqnarray}
Moreover the quantities $R(a_j,a_j)$ appearing in (\ref{dlog_tau})
are given by
\begin{equation}
 R(a_j,a_j)=\sum_{k\neq j}(-1)^k\,  {(q_j p_k - p_j q_k)^2\over a_j - a_k} \,
+ p_j^2 - a_j q_j^2 - 2 p_j q_j u + 2 q_j^2 v \, .\label{airy_7}
\end{equation}
These equations are derived very much in the spirit of
 \cite{tw1}; see also \cite{dyson92,its90,mehta92b}.
\subsection{The Ordinary Differential Equations}
\label{subsec:sum_ode}
  For the special case $J=(s,\infty)$
the above equations
 can be used
to show that $q(s;\lambda)$ (the quantity $q$ of the last section corresponding
to the end-point $s$)
 satisfies
\begin{equation}
q^{\prime\prime}= s\,  q + 2 q^3 \, ,\ \ \ (^\prime={d\over ds})
\label{p2}
\end{equation}
with
\begin{equation}
 q(s;\lambda) \sim \sqrt{\lambda}\> {\rm Ai}(s)
 \ \ \ {\rm as}\ \  s\rightarrow\infty.
\label{p2_bc}
\end{equation}
  This equation
is a special case of the ${\rm P}_{II}$ differential equation
\cite{ablowitz_segur,gauss_painleve,mtw,painleve}.
One can similarly derive for $R(s):=R(s,s)$, which in view of
(\ref{dlog_tau}) equals
\[ {d\over ds}\, \log D(J;\lambda)\, , \]
the third-order equation
\begin{equation}
{1\over 2}\left( {R^{\prime\prime}\over R^\prime}\right)^\prime
- {R\over R^\prime} + R^\prime = 0\, .
\label{rDE}
\end{equation}
It is also the case that
\begin{equation}
  R^\prime(s)=-q(s;\lambda)^2
\label{Rq_relation}
\end{equation}
 and this gives the following simple
formula for $D(J;\lambda)$
 in terms of a ${\rm P}_{II}$ transcendent:
\begin{equation}
 D(J;\lambda)  = \exp\left( -  \int_s^\infty  (x-s) q(x;\lambda)^2
  \, dx\right)  \, .
\label{det_p2}
\end{equation}
This is much simpler than the corresponding representation of
$D(J;\lambda)$
for the sine kernel in terms of a ${\rm P}_V$ transcendent.
The fact that $q(s;\lambda)$ satisfies (\ref{p2}) can also be
obtained by combining some results in
\cite{ablowitz_segur,clarkson_mcleod1,hastings_mcleod}.
Thus in this  case of a semi-infinite
interval  our results  connect with  inverse
scattering.
\subsection{Asymptotics and the  Commuting Differential Operator}
\label{subsec:sum_commuting}
Again we take $J=(s,\infty)$ and consider asymptotics as $s\rightarrow
-\infty$. (Asymptotics as $s\rightarrow\infty$ can be obtained
trivially from the Neumann series for $(I-K)^{-1}$.)\   From the
random matrix point of view the interesting quantities are
\[ E(n;s):= {(-1)^n\over n!}\> {\partial^n\over \partial \lambda^n}
\, D(I;\lambda)\biggr\vert_{\lambda=1} \, .\]
This is the probability that exactly $n$ eigenvalues lie in $J$.
Using our differential equations
plus the known asymptotics of $q(s;1)$ as $s\rightarrow -\infty$
\cite{hastings_mcleod} we obtain the asymptotic expansions
\begin{equation}
 R(s;1)\sim  {1\over 4} s^2 -{1\over 8\, s} + {9\over 64\,  s^4}
 -{189\over 128\,  s^7}
+ {21663\over 512\, s^{10}}  +\cdots \, \ \ \ (s\rightarrow -\infty) .
\label{R_exp}
\end{equation}
(where $R(s;1)$ denotes $R(s)$ corresponding to $\lambda=1$) and
\begin{eqnarray}
 E(0;s)& =& D(J;1) \sim  {\tau_0\over (-s)^{1/8}}\exp\left(s^3/12\right)
\nonumber \\
&&\times\left (1 -
{3\over 2^6\,  s^3} +{2025\over 2^{13}\, s^6}
- {2470825\over 2^{19}\, s^9}  + \cdots \right)
\ \ \ (s\rightarrow -\infty)
\label{E0_exp}
\end{eqnarray}
where $\tau_0$ is an undetermined constant.
\par
  The analogue  of this
formula for  the sine kernel  was  obtained by
Dyson \cite{dyson76}.
The analogue of our constant $\tau_0$ in the expansion was found
to be equal to $2^{1/12} e^{3\zeta^\prime(-1)}$ ($\zeta(s)$ is
the Riemann zeta function).  This constant was obtained by
scaling a result of the second author \cite{widom71} on Toeplitz
matrices.  Such a result is not available to us now.
Using (\ref{R_exp}) and the
fact that $R(s)$ is the logarithmic derivative
of $E(0;s)$,  it is straightforward to derive an integral
representation for $\log\tau_0$.  This integral can be  evaluated
approximately
by numerically integrating (\ref{p2}) and using (\ref{Rq_relation}).
This leads to the value $\log\tau_0\simeq -0.136540$.  We conjecture
that in fact
\[
\tau_0 = e^{\zeta^\prime(-1)}\,  2^{1\over 24} =
 0.87237\> 14149\>  54127\, \ldots\, .
\]
This agrees with the numerically computed value, up to its
degree of accuracy.
\par
For asymptotics of $E(n;s)$ for general $n$ we introduce
\[ r(n;s):= {E(n;s)\over E(0;s)}\, . \]
Successive differentiation of (\ref{p2}) with respect to $\lambda$,
plus the known asymptotics of $q(s;1)$, allows us to find asymptotic
expansions for the quantities
\begin{equation}
 q_n(s):= {\partial^n q\over \partial\lambda^n}\biggr\vert_{\lambda=1}
\label{qn_defn}
\end{equation}
(for the analogue  in the sine kernel case see \cite{btw}); and these
in turn can be used to find expansions for $r(n;s)$.  One drawback
of this approach is that yet another undetermined constant factor
enters the picture.  (In \cite{btw,widom92} Toeplitz and Wiener-Hopf
techniques,
not available for the Airy kernel, fixed this constant.)\ \   Another
drawback is that when one expresses the $r(n;s)$ in terms of the
$q_n(s)$ a large amount of cancellation takes place, with the result
that even the first-order asymptotics of $r(n;s)$ are out of reach
by this method when $n$ is large.
\par
There is, however, another approach (briefly indicated in \cite{btw},
and with details in \cite{tw1},  for the sine kernel case).
We have
\begin{equation}
r(n;s)=\sum_{i_1<\cdots<i_n} {\lambda_{i_1}\cdots \lambda_{i_n} \over
(1-\lambda_{i_1})\cdots (1-\lambda_{i_n}) }
\label{eig_repr}
\end{equation}
where $\lambda_0>\lambda_1>\cdots$ are the eigenvalues of the
integral operator $K$ (with $\lambda=1$).  Now just as
the operator with the sine kernel commutes with
the differential operator for the prolate spheroidal wave
functions \cite{ince},
so does  the Airy operator commutes
with the differential operator $L$ given by
\begin{equation}
 Lf(x)=\left(\left(x-s\right) f^\prime(x)\right)^\prime - x(x-s) f(x)\, .
\label{L_operator}
\end{equation}
An application of the WKB method, plus a trick
 allows us to
derive the following asymptotic formula for $\lambda_i$ with
$i$ fixed:
\begin{equation}
 1-\lambda_i\sim {\sqrt{\pi}\over i!}\,  2^{5i+3}\,  t^{3i/2 +3/4} \exp\left(
-{8\over 3} t^{3/2}\right) \ \ \ (s=-2t\rightarrow - \infty).
\label{eigs_asy}
\end{equation}
(The analogue of this for the sine kernel is in~\cite{fuchs64}.)
{}From this and  (\ref{eig_repr}) we deduce
\begin{equation}
r(n,-2t)\sim {1! 2! \cdots (n-1)!\over \pi^{n/2}\,  2^{(5n^2+n)/2}}\>
t^{-3n^2/4}\>\exp\left( {8\over 3} n t^{3/2} \right ) \, .
\label{r_asy}
\end{equation}
(Note that this can be used to fix the constant
mentioned in the last paragraph.)
\subsection{Probability Density  for the  $n^{th}$ Largest Eigenvalue}
An elementary probability argument shows that if $F(n;s)$
($n=0,1,\cdots $)  denotes
the probability density for the (scaled)  position
of the     $n^{th}$ largest eigenvalue
($n=0$ is  the largest), then
\[ F(0;s) = {d E(0;s)\over ds} \]
and for $n\geq 1$
\[ F(n;s)-F(n-1,s)={d E(n;s)\over ds}\, . \]
\par
{}From the preceding expressions  for $E(n;s)$ in terms of $q(s;1)$ and
$q_n(s)$ we can  derive formulas for $F(n;s)$ which can
in turn be numerically evaluated to produce tables of values
for these probability densities.  We have done this for
$n=0$ and $n=1$ and in
 Fig.~1 we plot $E(0;s)$ and $E(1;s)$  and in Fig.~2 we plot
$F(0;s)$ and $F(1;s)$.
\section{THE SYSTEM OF PARTIAL DIFFERENTIAL EQUATIONS}
\label{sec:pde}
\subsection{Derivation of the Equations}
\label{subsec:derivation_pde}
A central role in the derivation of our equations is played by the
commutator identity
\begin{equation}
\left\lbrack L,\left(I-K\right)^{-1}\right\rbrack = \left(I-K\right)^{-1}
\left\lbrack L,K\right\rbrack \left(I-K\right)^{-1}
\label{CR_1}
\end{equation}
for any operators $K$ and $L$. Taking $L$ to be $D$ ($=d/dx$) or
$M$ (=multiplication by the independent variable) give two of the three
basic ingredients from which the equations are derived.  The third
is the equally simple fact that if $K$ depends upon a parameter $\alpha$
then
\begin{equation}
{d\over d\alpha} \left(I-K\right)^{-1} = -\left(I-K\right)^{-1}
{dK\over d\alpha} \left(I-K\right)^{-1} \, .
\label{dResolvent_dalpha}
\end{equation}
\par
It will be very useful for us to think of $K$ as acting not on $J$
but on $(-\infty,\infty)$ and to have kernel
\begin{equation}
K(x,y) \chiJ (y)
\label{K_char}
\end{equation}
where $\chiJ$ is the characteristic function of $J$.  We continue
to denote the resolvent kernel of $K$ by $R(x,y)$ but note that
while it is smooth in $x$ it is discontinuous at $y=a_j$.  Thus the
quantity $R(a_j,a_j)$ appearing in (\ref{dlog_tau}) must be
interpreted to mean
\[ \lim_{y\rightarrow a_j \atop y\in J} R(a_j,y); \]
similarly for $p_j$ and $q_j$ in (\ref{qp_defn}) and like quantities
appearing later.  The definitions of $u$ and $v$ must be modified
to read
\begin{equation}
u=\left(A\chiJ,\left(I-K\right)^{-1}A\right),\ \ \
v=\left(A\chiJ,\left(I-K\right)^{-1}A^\prime\right)\, .
\label{uv_defn2}
\end{equation}
Notice that since
\[ \left(I-K\right)^{-1} A = \left(I-K\right)^{-1} A \chiJ \ \ \
{\rm in }\  J
\]
this agrees with the original  definitions of $u$ and $v$
given  by (\ref{uv_defn}).
\par
We can think of $K$ as an operator taking smooth functions to smooth
functions and so the operators $KD$ and $DK$ make sense, the former
having distributional kernel.  The transpose operator $K^t$ takes
distributions to distributions.
\par
With these preliminaries out of the way we begin with the
derivation of the representations of the quantities $R(a_j,a_k)$
in terms of the $q_j$, $p_j$, $u$ and $v$.  We introduce
the notation
\begin{equation}
Q=\left(I-K\right)^{-1} A, \ \ \ P=\left(I-K\right)^{-1} A^\prime \, ,
\label{QP_defn}
\end{equation}
so that $q_j=Q(a_j)$, $p_j=P(a_j)$.  (For the moment we ignore
the dependence of $Q$ and $P$ on the parameter $a=(a_1,\cdots,a_{2m})$;
this will come later.)
\par
The first commutator relation we use is the simple
\[ \left\lbrack M,K \right\rbrack  \doteq \left( A(x) A^\prime(y)
-A^\prime(x) A(y) \right) \chiJ (y) \]
(where $\doteq$ means ``has kernel equal to'').  Hence from (\ref{CR_1})
with $L=M$
\begin{equation}
\left\lbrack M,\left(I-K\right)^{-1}\right\rbrack\doteq
Q(x) \left(I-K^t\right)^{-1} A^\prime \chiJ (y) - P(x) \left(I-K\right)^{-1}
A\chiJ (y)
\label{CR_2}
\end{equation}
(The transpose here arises from the general fact that if $L\doteq U(x) V(y) $,
then $T_1L T_2\doteq T_1 U(x)\>  T_2^t V(y)$.)\ \ If we use the fact
that
\begin{equation}
\left(I-K^t\right)^{-1} A \chiJ = \left(I-K\right)^{-1} A
\ \ \ {\rm on} \ J,
\label{transpose_rela}
\end{equation}
we deduce from (\ref{QP_defn}) and (\ref{CR_2})  that \cite{its90}
\begin{equation}
R(x,y) = {Q(x) P(y) - P(x) Q(y) \over x-y }
 \ \ \ \left(x,y\in J, x\neq y\right).
\label{R_QP}
\end{equation}
In particular, therefore,
\begin{eqnarray}
R(a_j,a_k)={q_j p_k - p_j q_k\over a_j-a_k} \ \ \ (j\neq k),
 \label{R_jk}  \\
R(a_j,a_j)=Q^\prime(a_j) p_j - P^\prime(a_j) q_j \ \ \
\left(^\prime={d\over dx}\right)\, .
\label{R_jj}
\end{eqnarray}
\par
To compute $Q^\prime(x)$ and $P^\prime(x)$ we consider first the
commutator $[D,K]$.  If an operator $L$ has distributional kernel
$L(x,y)$ then
\begin{equation}
\left\lbrack D,L\right\rbrack \doteq \left( {\partial\over\partial x}+
{\partial\over\partial y}\right) L(x,y)\, .
\label{CR_3}
\end{equation}
It is an easy verification, using the equation $A^{\prime\prime}(x)=x A(x)$,
that
\begin{equation}
\left( {\partial\over\partial x}+
{\partial\over\partial y}\right) K(x,y) = - A(x) A(y)
\label{partial_K}
\end{equation}
and so (recall the definition (\ref{K_char}) of the kernel of $K$)
\[
\left\lbrack D,K\right\rbrack \doteq - A(x)A(y)\chiJ (y)
-\sum_k (-1)^k K(x,a_k)\delta(y-a_k)\, .
\]
Hence, applying (\ref{CR_1}) with $L=D$ (and recalling definitions
(\ref{QP_defn})),
\begin{equation}
\left\lbrack D,\left(I-K\right)^{-1}\right\rbrack
=-Q(x) \left(I-K^t\right) A \chiJ (y) - \sum_k(-1)^k R(x,a_k)\rho(a_k,y)
\label{CR_5}
\end{equation}
where
\[ \rho(x,y)=\delta(x-y)+R(x,y)\]
is the distributional kernel of $(I-K)^{-1}$.
\par
We proceed with the computations of $Q^\prime(x)$ and $P^\prime(x)$.
Differentiating the first relation in (\ref{QP_defn}) and
using (\ref{CR_5}) give
\begin{eqnarray*}
Q^\prime&=& D(I-K)^{-1} A = P+\left\lbrack D,\left(I-K\right)^{-1}
\right\rbrack A \\
&=& P - Q\left(\left(I-K^t\right)^{-1}A\chiJ,A\right) -
\sum_k(-1)^k R(\cdot,a_k) q_k \, .
\end{eqnarray*}
Thus (recall (\ref{uv_defn2}))
\begin{equation}
Q^\prime(x)=P(x) - Q(x) u - \sum_k(-1)^k R(x,a_k) q_k\, .
\label{Q_deriv}
\end{equation}
\par
Similarly
\[ P^\prime = (I-K)^{-1} A^{\prime\prime} +
 \left\lbrack D,\left(I-K\right)^{-1}\right\rbrack A^\prime\, . \]
Since $A^{\prime\prime}(x)=xA(x)$ this is equal to
\[ M\left(I-K\right)^{-1} A - \left\lbrack M,\left(I-K\right)^{-1}\right
\rbrack A +\left\lbrack D,\left(I-K\right)^{-1}\right\rbrack A^\prime\, . \]
We now use formula (\ref{CR_2}) and (\ref{CR_5}) to deduce that
\begin{equation}
P^\prime(x)=x Q(x) - 2 Q(x) v + P(x) u - \sum_k (-1)^k R(x,a_k) p_k \, .
\label{P_deriv}
\end{equation}
Setting $x=a_j$ in ({\ref{Q_deriv}) and (\ref{P_deriv}) and using
(\ref{R_jj}) give
\[ R(a_j,a_j)=p_j^2-a_j q_j^2 +2 q_j^2 v - 2 p_j q_j u
+\sum_k (-1)^k R(a_j,a_k) (q_j p_k - p_j q_k)\, .
\]
In view of (\ref{R_jk}) this is precisely equation (\ref{airy_7}).
\par
Most of the work is already done and we can derive the equations
(\ref{airy_1})--(\ref{airy_6}) very quickly.  First, we have the
easy fact
\[ {\partial\over\partial a_k} K\doteq (-1)^k K(x,a_k)\delta(y-a_k) \]
and so by (\ref{dResolvent_dalpha})
\begin{equation}
{\partial\over \partial a_k} \left(I-K\right)^{-1} \doteq
(-1)^k R(x,a_k) \rho(y-a_k) \, .
\label{dR2}
\end{equation}
At this point we must keep in mind that $Q$ and $P$ are functions
of $a$ as well, and so we denote them now by $Q(x,a)$ and
$P(x,a)$, respectively.  We deduce immediately from (\ref{dR2})
and the definitions (\ref{QP_defn}) that
\begin{equation}
{\partial\over\partial a_k} Q(x,a) = (-1)^k R(x,a_k) q_k, \ \ \
{\partial\over\partial a_k} P(x,a) = (-1)^k R(x,a_k) p_k\, .
\label{dQP}
\end{equation}
Since $q_j=Q(a_j,a)$ and $p_j=P(a_j,a)$ this gives
\[ {\partial q_j\over \partial a_k}= (-1)^k R(a_j,a_k) q_k ,\ \ \
{\partial p_j\over \partial a_k}= (-1)^k R(a_j,a_k) p_k \ \ \ (j\neq k).\]
In view of (\ref{R_jk}) these are equations (\ref{airy_1}) and
(\ref{airy_2}).  Moreover
\[ {\partial q_j\over \partial a_j}=\left( {\partial\over \partial x}+
{\partial\over \partial a_j}\right)Q(x,a)\biggr\vert_{x=a_j}\ ,
\ \ \ \
 {\partial p_j\over \partial a_j}=\left( {\partial\over \partial x}+
{\partial\over \partial a_j}\right)P(x,a)\biggr\vert_{x=a_j} \]
and (\ref{dQP}), (\ref{Q_deriv}) and (\ref{P_deriv}) give
\begin{eqnarray*}
{\partial q_j\over \partial a_j}&=& p_j - q_j u -
 \sum_{k\neq j} (-1)^k R(a_j,a_k)
q_k\, ,  \\
{\partial p_j\over \partial a_j}&=& a_jq_j-2q_j v +p_j u - \sum_{k\neq j}
(-1)^k R(a_j,a_k) p_k \, .
\end{eqnarray*}
In view of (\ref{R_jk}) again, these are equations (\ref{airy_3})
and (\ref{airy_4}).
\par
Finally, using the definition of $u$ in (\ref{uv_defn2}), the fact
\[ {\partial\over \partial a_j}\chiJ (y)=(-1)^j\delta(y-a_j)\, ,\]
and (\ref{dR2}) we find that
\[ {\partial u\over \partial a_j}=(-1)^j A(a_j) q_j +
(-1)^j \left(A\chiJ,R(\cdot,a_j)\right) q_j\, .\]
But
\[\left(A\chiJ,R(\cdot,a_j)\right) = \int_J R(x,a_j) A(x)\, dx
=\int_J R(a_j,x) A(x)\, dx \]
since $R(x,y)=R(y,x)$ for $x,y\in J$.  Since $R(y,x)=0$ for
$x\not\in J$ the last integral equals
\[ \int_{-\infty}^\infty R(a_j,x) A(x)\, dx = A(a_j) - q_j\, . \]
Thus
\[ {\partial u\over \partial a_j}= (-1)^j q_j^2\, ,\]
which is (\ref{airy_5}).  Equation (\ref{airy_6}) is derived
analogously.
\par
We end this section with two relations between $u$, $v$, the $q_j$
and the $p_j$ which in fact can be used to represent $u$ and $v$
directly in terms of the $q_j$ and $p_j$.  These relations are
\begin{eqnarray}
2v - u^2 &= &\sum_j (-1)^j q_j^2\, , \label{uv_1}\\
u&=& -\sum_j\left(p_j^2-a_j q_j^2-2p_jq_j u +2q_j^2 v\right) \, .
\label{uv_2}
\end{eqnarray}
\par
To obtain the first of these observe that (\ref{airy_1}) and
(\ref{airy_3}) imply
\begin{equation}
\left(\sum_k{\partial\over\partial a_k}\right) q_j = p_j-q_j u
\label{uv_3}
\end{equation}
while from (\ref{airy_5}) and (\ref{airy_6})
\[
{\partial\over\partial a_j}\left(2v-u^2\right)=
2(-1)^j q_j\left(p_j-q_j u\right)\, .
\]
Hence if we multiply both sides of (\ref{uv_3}) by $2(-1)^j q_j$
and sum over $j$ we obtain
\[
\left(\sum_k {\partial\over\partial a_k}\right) \left(
\sum_j (-1)^j q_j^2\right) = \left(\sum_k {\partial\over\partial a_k}\right)
\left(2v-u^2\right)\, .
 \]
It follows that the two sides of (\ref{uv_1}) differ by a function
of $(a_1,\cdots,a_{2m})$ which is invariant under translation by
any vector $(s,\cdots,s)$.  Since, clearly, both sides tend to zero
as all $a_i\rightarrow\infty$, their difference must be identically
zero.
\par
To deduce (\ref{uv_2}) we add
(\ref{CR_5}) and (\ref{dR2}) to obtain
\[ \left( {\partial\over\partial x}+{\partial\over\partial y}+
\sum_k {\partial\over\partial a_k}\right) R(x,y) = -Q(x) Q(y)
\ \ \ (x,y\in J).\]
(Observe that the kernel of $[D,(I-K)^{-1}]$ is
$(\partial/\partial x + \partial/\partial y ) R(x,y)$.)\ \
Hence
\[ \left(\sum_k {\partial\over\partial a_k}\right) R(a_j,a_j)
= -q_j^2=-(-1)^j {\partial u\over\partial a_j}\, , \]
the last by (\ref{airy_5}).  Multiplying by $(-1)^j$ and summing
over $j$ gives
\[ \left(\sum_k {\partial\over\partial a_k}\right)
\left(\sum_j(-1)^j R(a_j,a_j)\right) = -\left(\sum_k {\partial\over
\partial a_k}\right) u\, .\]
{}From this we deduce, by an argument similar to the one at the
end of the last paragraph, that
\begin{equation}
\sum_j(-1)^j R(a_j,a_j) = -u\, .
\label{uv_5}
\end{equation}
If we substitute formula (\ref{airy_7}) into this we see that the
resulting double sum vanishes and we are left with (\ref{uv_2}).
\subsection{Hamiltonian Structure}
\label{subsec:hamiltonian}
The partial differential equations (\ref{airy_1})--(\ref{airy_6}) can
be rewritten in a compact form  that
 reveals  a symplectic structure with
(\ref{airy_7}) defining a family of  commuting Hamiltonians.
  To see this we first
change the notation slightly
\begin{eqnarray*}
q_{2j}&=& -{i\over 2} x_{2j},\ \ \ \ p_{2j}=-i y_{2j}, \\
q_{2j-1}&=& {1\over 2} x_{2j-1},\ \ \ \ p_{2j-1}=y_{2j-1},
\end{eqnarray*}
for $j=1,2,\ldots,m$, and
\[ v={1\over 2} x_0,  \ \ \ \ u=y_0,  \]
and introduce the canonical symplectic structure
\[ \lbrace x_j,x_k\rbrace = \lbrace y_j,y_k\rbrace  = 0, \ \ \
\lbrace x_j,y_k\rbrace = \delta_{jk} ,\]
for $j,k=0,1,\ldots,m$. Then equations (\ref{airy_1})--(\ref{airy_6})
become
\begin{equation}
d_a x_j = \lbrace x_j,\omega(a)\rbrace \ \ \  {\rm and} \ \ \
d_a y_j=\lbrace y_j,\omega(a)\rbrace \ \ \ (j=0,1,\ldots,m)
\label{hamilton_eqn}
\end{equation}
where
\begin{eqnarray}
\omega(a) &=&  d_a \log {\rm det}\left(I-K\right) \nonumber \\
&=& \sum_{j=1}^{2m} G_j(x,y)\, da_j
\label{hamiltonian}
\end{eqnarray}
and
\[
G_j(x,y) = y_j^2 - {1\over 4} a_j x_j^2 - x_j y_j y_0 +{1\over 4} x_j^2 x_0
-{1\over 4}\sum_{k=1\atop k\neq j}^{2m}
{\left(x_j y_k - x_k y_j\right)^2\over a_j - a_k} \> .
\]
Furthermore, the $G_j$'s are in involution
\[
\left\lbrace G_j,G_k \right\rbrace =0  \, .
\]
This last result can be verified by a direct calculation, but as
we discuss below, there is a better way to understand why such
Hamiltonians are in involution.
We can summarize the system of equations by
saying, in words, that
  to find the $a_k$-flow of the coordinates  $x_j$ and  $y_j$,
one flows according to Hamilton's equation with Hamiltonian $G_k$.
The consistency of
these equations  (i.e.\ mixed partials are equal) follows immediately
from this Hamiltonian formulation (see, e.g.\  \cite{tw1}).
\par
When equations (\ref{airy_1})--(\ref{airy_6}) are expressed
in the Hamiltonian form (\ref{hamilton_eqn}), the equations are
exactly analogous to the case of the sine kernel \cite{jmms} and
in both cases
the Hamiltonians are  defined via (\ref{hamiltonian}).
For the JMMS equations (see, e.g.\   \cite{tw1})
this Hamiltonian system is closely related to  the Hamiltonian
system  studied
by Moser \cite{moser}.  In fact in both cases
 the  systems of
partial differential equations  can be
formulated in terms of the    classical $R$-matrix approach to
completely integrable systems.  Such an approach identitfies
the Lax operator and hence  the commuting flow result above
follows without any calculations.
  We refer the reader to \cite{htw}
for a discussion of this classical $R$-matrix  formulation
of these systems of partial differential equations.

\section{THE ORDINARY DIFFERENTIAL EQUATIONS}
\label{sec:ode}
In this section we specialize to the case $J=(s,\infty)$ and derive
the differential equations (\ref{p2}), (\ref{rDE}) and (\ref{Rq_relation}).
In the notation of the last section $m=1$, $a_1=s$, $a_2=\infty$.
All terms in the formulas we derived which contain $a_2$ vanish
because of the decay of the Airy kernel at $+\infty$.  Recall
that we now denote the quantity $q$ corresponding to the endpoint
$a_1=s$ by $q(s)$, and write $R(s)$ for $R(s,s)$.
\par
First, equation (\ref{Rq_relation}) is immediate from (\ref{uv_5}),
which now says that $R=u$, and (\ref{airy_5}).
\par
For (\ref{rDE}) we observe that (\ref{airy_3}) and (\ref{airy_4}) in
the present case read
\begin{eqnarray}
q^\prime &=& p-q u\, ,\label{q_prime}\\
p^\prime &=& s q +p u - 2 q v\, , \label{p_prime}
\end{eqnarray}
whereas (\ref{airy_7}) reads
\[ R=p q^\prime - q p^\prime \, .  \]
Thus, by (\ref{Rq_relation}),
\begin{equation}
{R\over R^\prime} = \left({p\over q}\right)^\prime\, .
\label{Rprime}
\end{equation}
Taking the logarithmic derivative of both sides of (\ref{Rq_relation})
gives
\[ {1\over 2}{R^{\prime\prime}\over R^\prime} = {q^\prime\over q} \]
and by (\ref{q_prime}) this equals
\[ {p\over q}-u={p\over q} - R \, . \]
Thus
\[ {1\over 2} \left({R^{\prime\prime}\over R^\prime}\right)^\prime
= \left( p\over q \right)^\prime - R^\prime \, . \]
By (\ref{Rprime}) this is just (\ref{rDE}).
\par
Finally, to obtain (\ref{p2}) we differentiate (\ref{q_prime})
and use (\ref{q_prime}), (\ref{p_prime}), (\ref{airy_5}) and
(\ref{airy_6}).  We find by an easy computation that
\[ q^{\prime\prime} = s q +q^3 +q(u^2-2v)\, . \]
Conveniently, we have (\ref{uv_1}), which says in this case
that $u^2-2v=q^2$.  Thus (\ref{p2}) is established.
\section{ASYMPTOTICS}
\label{sec:asy}
\subsection{Asymptotics via  Painlev{\'e}}
\label{subsec:asy_painleve}
Hastings and McLeod \cite{hastings_mcleod}
(see also \cite{clarkson_mcleod1,clarkson_mcleod2})
 have shown that $q(s;1)$, the unique
solution to equation (\ref{p2}) which is asymptotic to ${\rm Ai}(s)$
as $s\rightarrow +\infty$, is asymptotically equal to $\sqrt{-s/2}$
as $s\rightarrow -\infty$.
In fact there is  a complete asymptotic expansion in decreasing
powers of $-s$,
beginnning with this term \cite{mcleod_private}.  Successive terms
are easily computed.
This result
and later ones, become a little easier to state if we write
$s=-t/2$:
\begin{equation}
q(-t/2;1)={1\over 2}\sqrt{t}\left( 1 - {1\over t^3} - {73\over 2t^6}
-{10657\over 2 t^9} - {13912277\over 8 t^{12}}
+{\rm O}({1\over t^{15}})\right) \ \ \
(t\rightarrow +\infty)
\label{p2_asy}
\end{equation}
Squaring and integrating and using (\ref{Rq_relation}) we deduce
\[ R(s) = {1\over 4} s^2 + c - {1\over 8\, s} + {9\over 64\,  s^4} -
{189\over 128 \, s^7} + {21663\over 512\,  s^{10}} + {\rm O}({1\over s^{13}})
\ \ \ (s\rightarrow -\infty)  \]
where $c$ is a constant.  Substituting this into (\ref{rDE}) shows
that $c=0$.  Thus we obtain (\ref{R_exp}) and then, after integrating
and exponentiating, (\ref{E0_exp}).
\par
To obtain the asymptotic expansions of $q_n(s)$ in (\ref{qn_defn}),
we differentiate (\ref{p2}) with respect to $\lambda$ and set
$\lambda=1$.  We obtain
\[ q_1^{\prime\prime}-\left(s+6q_0^2\right) q_1=0 \]
(where $q_0(s)=q(s;1)$).  Using (\ref{p2_asy}) we find two linearly
independent solutions of this equation, one exponentially large at
$-\infty$ and one exponentially small.  Assuming the exponentially
large one actually appears with a nonzero factor we obtain the
asymptotic expansion
\begin{equation}
q_1(-t/2)=c_1 {\exp({1\over 3} t^{3/2})\over t^{1/4}}
\left( 1 + {17\over 24\,  t^{3/2}} +{1513\over 2^7 3^2\,  t^3} +
{850193\over 2^{10} 3^4\,  t^{9/2}}+{407117521\over 2^{15} 3^5\,  t^{6}}
 + \cdots
\right)
\label{q1_asy}
\end{equation}
where $c_1$ is a constant factor to be determined.
\par
Successive differentiations of (\ref{p2}) with respect to
$\lambda$, followed by setting $\lambda=1$, yields a sequence
of linear differential equations for the $q_n$,
\[ q_n^{\prime\prime}-\left(s+6q_0^2\right) q_n= \varphi_n(q_0,\cdots,
q_{n-1}),
\]
where $\varphi_n$ is a polynomial in the $q_m$ with $m<n$.  For $n>1$
a particular solution dominates all solutions of the homogeneous
equation.  Therefore, in the asymptotics, no new unknown constant
factors are introduced (unless we want to go ``beyond all orders'').
\par
To deduce expansions for the $r(n;s)$ we must differentiate (\ref{det_p2})
$n$ times with respect to $\lambda$ and set $\lambda=1$.  We find
expressions of the form
\begin{equation}
r(n;s)=\int_s^\infty (x-s) \psi(q_0,\cdots,q_n)\, dx
\label{r_integral}
\end{equation}
where $\psi_n$ is a polynomial in $q_0, \cdots, q_n$.  In particular
we obtain (after setting $s=-t/2$)
\[ r(1;-t/2)={1\over 2}\int_{-\infty}^t (t-x) q_0(-x/2) q_1(-x/2)\, dx\, . \]
Substituting (\ref{p2_asy}) and (\ref{q1_asy}) into this  integral
gives the asymptotic expansion for $r(1;-t/2)$.  The first order
result is
\[ r(1;-t/2) \sim c_1 {\exp(t^{3/2}/3)\over t^{3/4}} \]
and comparing this with (\ref{r_asy}) reveals that
\[ c_1={1\over 2\sqrt{2\pi}}\, .\]
\par
We now state the expansions we have obtained
in this way  with the help of
Mathematica.  For $q_n$ with $n>2$, as $t\rightarrow +\infty$
\begin{eqnarray}
q_n(-t/2)&=& {n!\over 2^{3n/2}\, \pi^{n/2}}\,
{ \exp({n\over 3} t^{3/2}) \over
t^{3n/4-1/2}}
\Biggl( 1 +{59n-36\over 24}\>  t^{-3/2} + \nonumber \\
&& \quad {3481 n^2+8244 n -8496\over 1152 }\>  t^{-3} +
 {\rm O}(t^{-9/2})\Biggr).
\label{qn_asy}
\end{eqnarray}
For $n=1, 2$ the leading order term is as above.  For $n=1$ the
correction terms are given above in (\ref{q1_asy}).  For $n=2$ the
coefficient of $t^{-3/2}$ is as given above but the coefficient
of $t^{-3}$ is\
$ {5461/ 288}$.
(Similar exceptional values of $n$ occurred in analogous expansions
for the sine kernel \cite{btw}.)\ \ These expansions have been
computed for $n\leq 10$ and are undoubtedly correct for all $n$.
(In \cite{btw} the analogous expansions were proved for all $n$ by
a backwards induction argument---such arguments should extend to the
present case of $q_n$.)
\par
For $r(n;s)$ we have obtained the following sharpening of
(\ref{r_asy}): as $t\rightarrow +\infty$
\begin{eqnarray}
r(n;-t/2)& =& {1! 2! \cdots (n-1)!\over 2^{n^2+n/2} \pi^{n/2} } \,
{\exp({n\over 3} t^{3/2})\over t^{3n^2/4}}
 \Biggl(
1+
{n(34 n^2 +31)\over 24}\>  t^{-3/2} + \nonumber \\
&& \quad {n^2(1156 n^4+6608 n^2 +11509)\over 1152 }\>  t^{-3}+
{\rm O}(t^{-9/2}) \Biggr).
\label{r_asy2}
\end{eqnarray}
This was established for $n\leq 6$.  As was mentioned in the Introduction,
when we substitute expansions (\ref{qn_asy}) into (\ref{r_integral})
a large amount of cancellation takes place, so that $r(n;s)$ is
much smaller than might be anticipated.  This also means that to
deduce even the first-order asymptotics in (\ref{r_asy2})
for moderate values of $n$ one has to go very far out in the
expansions (\ref{qn_asy}).  Thus it is important that we have an
alternative approach which yields asymptotic results for all values
of $n$.

\subsection{Asymptotics via the Commuting Differential Operator}
\label{subsec:asy_diffop}
We begin with the integral identity
\begin{equation}
 K(x,y)=\int_0^\infty A(x+z) A(y+z)\, dz \, .
\label{K_identity}
\end{equation}
This appears  in \cite{clarkson_mcleod1}
 but we give a simple proof here.
Think of (\ref{partial_K}) as a differential equation for the ``unknown''
function $K(x,y)$.  Since
\[ \left({\partial\over\partial x}+{\partial\over \partial y}\right)
A(x+z) A(y+z) = {\partial\over\partial z} A(x+z) A(y+z) \]
one solution is the right side of (\ref{K_identity}) and, of course,
another solution is the Airy kernel $K(x,y)$.  The difference is
therefore of the form $\varphi(x-y)$ for some function $\varphi$.
Since both sides of (\ref{K_identity}) tend to zero as $x$ and $y$
tend to $+\infty$ independently, we must have $\varphi\equiv 0$.
\par
Rewriting (\ref{K_identity}) as
\[ K(x,y) = \int_s^\infty A(x+z-s) A(z+y-s) \, dz \]
shows that $K$ as an operator on $(s,\infty)$ is the square of the
operator with kernel $A(x+y-s)$.
\par
Integration by parts shows that an integral operator with kernel
$L(x,y)$ on $(s,\infty)$ commutes with a differential operator
\[ {d\over dx} \alpha(x) {d\over dx} + \beta(x) \]
if $\alpha(s)=0$ and if
\[ \alpha(y) {\partial^2L\over \partial y^2} + \alpha^\prime(y)
{\partial L\over \partial y} + \beta(y) L =
\alpha(x) {\partial^2 L\over \partial x^2} + \alpha^\prime(x)
{\partial L\over \partial x}+\beta(x) L \, .\]
(Of course we also require, in the end, appropriate vanishing at
$\infty$.)\ \ If we set $L(x,y)=A(x+y-s)$ and use the fact
$A^{\prime\prime}(x)=xA(x)$ we can easily check that a solution is given by
\[ \alpha(x)=x-s, \ \ \ \beta(x)=-x(x-s) \]
and so we obtain the commuting differential operator (\ref{L_operator}).
\par
Here is an outline of how (\ref{eigs_asy}) is found.  If we let
$f_1$, $f_2, \cdots $ denote the eigenfunctions of $L$ then the
logarithmic derivative of each $\lambda_i$ (thought of as a function
of $s$) is expressible very simply in terms of the corresponding
$f_i$.  The asymptotics of these eigenfunctions, and so of the
logarithmic derivative, is established by WKB techniques.  Integrating,
and using the fact that each $\lambda_i\rightarrow 1$ as
$s\rightarrow -\infty$,  gives (\ref{eigs_asy}).
The WKB argument we give is quite heuristic although it could
very likely be made rigorous.
\par
We make a preliminary change of variables, replacing $x$ by $x+s$
(so that $L$ acts on $(0,\infty)$) and $s$ by $-2t$.  The eigenvalue
problem can then be written as
\begin{equation}
\left(x f^\prime(x)\right)^\prime +\left(\mu-(x-t)^2\right)f(x)=0
\label{eig_eqn1}
\end{equation}
or in the equivalent form
\begin{equation}
x g^{\prime\prime}(x) +\left\lbrack \mu - (x-t)^2 +{1\over 4x}
\right\rbrack g(x)=0\, , \ \ \ \left(g(x)=\sqrt{x} f(x)\right)
\label{eig_eqn2}
\end{equation}
The eigenvalues satisfy $0<\mu_0<\mu_1<\cdots $ and oscillation
considerations applied to (\ref{eig_eqn2}) show that
$\mu_i={\rm O}(t^{1/2})$, for each $i$ as $t\rightarrow\infty$.
So we think of $\mu$ in our equations as ${\rm O}(t^{1/2})$, we
normalize our eigenfunctions so that $f(0)=1$, and proceed to
find asymptotics for large $t$.
\subsubsection{The region $x<<t^{-1/2}$}
\label{subsubsec:region1}
If we make the substitutions
\[ f(x) = h(t^2x)=h(y) \]
then (\ref{eig_eqn1}) becomes
\[ \left(y h^\prime(y) \right)^\prime - h(y) = t^{-2}\left(
{y^2\over t^4} -2 {y\over t} -\mu\right) h(y)\, .\]
The solution of the equation with the right hand side replaced  by $0$
which satisfies $h(0)=1$ is $I_0(2\sqrt{y})$ where $I_0$ is the
usual modified Bessel function.  The actual solution $h(y)$ will be
asymptotic to this in any interval over which the integral of the
factor of $h(y)$ on the right hand side of the equation is
${\rm o}(1)$.  Recalling that $\mu={\rm O}(t^{1/2})$ we see that
$y<<t^{3/2}$ suffices.  Hence
\begin{equation}
f(x)\sim I_0\left(2t\sqrt{x}\right)\, , \ \ \ (x<<t^{-1/2})
\label{appx1}
\end{equation}
\subsubsection{ The region $x>>t^{-2}$, $t-x>>t^{1/4}$}
\label{subsubsec:region2}
If in equation (\ref{eig_eqn2}) we set
\[ \alpha(x)={(t-x)^2-\mu\over x}, \ \ \  y=\int_0^x\sqrt{\alpha(z)}\, dz\, ,
\ \ \ h(y)=\alpha(x)^{1/4} g(x) \]
then the equation for $h(y)$ is
\[ h^{\prime\prime}(y)-h(y)=\left\lbrack {1\over 4} {\alpha^{\prime\prime}\over
\alpha^2} - {5\over 16} {{\alpha^\prime}^2\over \alpha^3}
-{1\over 4 x^2  \alpha}
\right\rbrack h(y)\, ,\]
where on the right side $^\prime= {d\over dx}$.  Since
$dy=\alpha(x)^{1/2}\, dx$, the integral with respect to $y$ (of the
factor of $h(y)$ on the right side of the equation) over any interval
equals
\[ \int \left( {1\over 4} {\alpha^{\prime\prime}\over \alpha^{3/2}}
-{5\over 16}{{\alpha^\prime}^2\over\alpha^{5/2}}-
{1\over 4 x^2 \alpha^{1/2}}\right)\, dx \]
over the corresponding $x$ interval.
This will be ${\rm o}(1)$ as long as, in this interval, $x>>t^{-2}$
and $t-x>>t^{1/4}$.
\par
In this same region $y>>1$ and so
\begin{equation}
h(y)\sim a e^y
\label{appx2}
\end{equation}
for some constant $a=a(t)$.  In case $x<<1$ we have $y=2t\sqrt{x} +
{\rm o}(1)$ and so
\[ f(x)=x^{-1/2} g(x)\sim a t^{-1/2} x^{-1/4} e^{2t\sqrt{x}}\, ,
\ \ \ (t^{-2}<<x<<1). \]
Comparing this result with (\ref{appx1}) in the overlapping region
$t^{-2}<<x<<t^{-1/2}$ and using the known asymptotics of $I_0(z)$
as $z\rightarrow\infty$ shows that $a\sim 1/2\sqrt{\pi}$.
Hence from (\ref{appx2})
\[ f(x)\sim {1\over 2\sqrt{\pi}} {(t-x)^{-1/2}\over x^{1/4}}
\exp\left\lbrace\int_0^x\sqrt{{(t-z)^2-\mu\over z}}\, dz\right\rbrace\, ,
\ \  (x>>t^{-2}, t-x>>t^{1/4})\, . \]
We also used here once again the estimate $\mu={\rm O}(t^{1/2})$.
Because of this same estimate we can write the integral in the
exponential as
\[ \int_0^x {t-z\over\sqrt{z}}\left\lbrace 1 - {1\over 2}
{\mu\over (t-z)^2}+{\rm O}\left({\mu^2\over (t-z)^4}\right)\right\rbrace
\, dz\, . \]
Because $t-x>>t^{1/4}$ the last term in brackets contributes ${\rm o}(1)$
to the integral, and so we obtain
\[ 2t x^{1/2} -{2\over 3} x^{3/2} -{\mu\over 2}\int_0^x {dz\over
\sqrt{z} (t-z)} \, . \]
Replacing $z$ by $tz^2$ shows that the integral here equals
\[ t^{-1/2} \log {\left(\sqrt{t}+\sqrt{x}\right)^2\over t-x} \, . \]
Thus
\begin{eqnarray}
 f(x)&\sim &  {1\over 2\sqrt{\pi}}{\left(t-x\right)^{{1\over 2} t^{-1/2}\mu -
{1\over 2}}\over x^{1/4} \left(\sqrt{t}+\sqrt{x}\right)^{
t^{-1/2}\, \mu}}\, \exp\left\lbrace 2 t x^{1/2} - {2\over 3} x^{3/2}\right
\rbrace
\label{appx3}\\
&&\hspace{3in} (x>>t^{-2}, t-x>>t^{1/4}).
\nonumber
\end{eqnarray}
\par
Let us see what happens if $x$ is not far from $t$, if
\[ t^{1/4}<<t-x<<t^{1/2} \, . \]
Then since
\[ 2t x^{1/2} -{2\over 3} x^{3/2} = {4\over 3} t^{4/3} - {1\over 2}
(x-t)^2 + {\rm o}(1) \]
in this region, we find
\begin{eqnarray}
f(x)&\sim & {1\over 2\sqrt{\pi}} 2^{-t^{-1/2}\mu} t^{-3/4}
\left( t\over t-x\right)^{-{1\over 2} t^{-1/2}\mu+{1\over 2}}
\exp\left\lbrace {4\over 3} t^{4/3}-{1\over 2} (x-t)^2\right\rbrace
\label{appx4}\\
&&\hspace{3in} (t^{1/4}<<t-x<<t^{1/2})
\nonumber
\end{eqnarray}
\subsubsection{The region $x-t>>t^{1/4}$}
\label{subsubsec:region3}
This is quite similar to the preceding.  The main difference
is that instead of making the variable change $y=\int \alpha(x)^{1/2}\, dx$
in (\ref{eig_eqn2}) we now set
\[ y={2\over 3} x^{3/2} -2 t x^{1/2} - \int_x^\infty
\left\lbrack \sqrt{\alpha(z)} - z^{1/2} + t z^{-1/2}\right\rbrack \, dz\, .
\]
Since an eigenfunction cannot grow exponentially at $\infty$ only
the negative exponential $e^{-y}$ can appear in the asymptotics and we
find that
\[ f(x)\sim b {(x-t)^{-1/2}\over x^{1/4}} \exp\left\lbrace
2 t x^{1/2} - {2\over 3} x^{3/2} + \int_x^\infty
\left\lbrack \sqrt{{(z-t)^2-\mu\over z}}-z^{1/2} + t z^{-1/2}\right\rbrack
\, dz\right\rbrace
\]
for some constant $b=b(t)$.  Now the integral in the exponential is
\[ -{\mu\over 2} t^{-1/2} \log{\left(\sqrt{x}+\sqrt{t}\right)^2\over
x-t} + {\rm o}(1) \]
and so
\begin{equation}
f(x)\sim b {\left(x-t\right)^{{1\over 2}t^{-1/2}\mu-{1\over 2}}\over
x^{1/4} \left(\sqrt{x}+\sqrt{t}\right)^{t^{-1/2}\mu}}
\exp\left\lbrace 2 t x^{1/2} - {2\over 3} x^{3/2}\right\rbrace
\ \ \ (x-t>>t^{1/4})
\label{appx5}
\end{equation}
and specializing to the region $t^{1/4}<<x-t<<t^{1/2}$ gives
\begin{eqnarray}
f(x)&\sim& b\,  2^{-t^{-1/2}\mu} t^{-3/4} \left({t\over x-t}\right)^{
-{1\over 2} t^{-1/2}\mu +{1\over 2}}\exp\left\lbrace
{4\over 3} t^{3/2} - {1\over 2} (x-t)^2\right\rbrace\label{appx6} \\
&&\hspace{3in}  (t^{1/4}<<x-t<<t^{1/2}) \nonumber
\end{eqnarray}
\subsubsection{The region $\vert x-t \vert <<t$}
\label{subsubsec:region4}
If in equation (\ref{eig_eqn2}) we set
\[ g(x)=h\left(\sqrt{2} t^{-1/4}(t-x)\right)=h(y) \]
and write $\varepsilon = (\sqrt{2} t^{3/4})^{-1}$ then the equation
becomes
\[ \left(1-\varepsilon y\right) h^{\prime\prime}(y)
+\left\lbrack {\mu\over 2 t^{1/2}} - {y^2\over 4} + {\varepsilon^2\over
4(1-\varepsilon y)}\right\rbrack h(y)=0\, . \]
Now we expect that for bounded $y$, and therefore also if $\vert y \vert
\rightarrow\infty$ at some rate (depending upon $\varepsilon$),
the solution of this will be asymptotic to a solution of
\[ h^{\prime\prime}(y) +\left( {\mu\over 2 t^{1/2}}-{y^2\over 4}\right)
h(y)=0\, , \]
Weber's equation.  Now unless
\[ {\mu\over 2 t^{1/2}} = i + {1\over 2} \ \ \ (i=0,1,\cdots) \]
there is no solution which vanishes at both $\pm\infty$.
But (\ref{appx4}) and (\ref{appx6}) show that in their regions
of validity (which overlap with our present region) $f(x)$ is a factor
depending on  $t$ times a factor tending exponentially to $0$
as $x-t\rightarrow\pm\infty$.  Hence our eigenvalues must satisfy
\[ {\mu_i\over 2 t^{1/2}}= i +{1\over 2} + {\rm o}(1) \]
and $h(y)$ is asymptotically a constant times the parabolic cylinder
function $D_i(y)$ (see, e.g.\ \cite{erdelyi}, Chp.~8).
  Thus (since $x\sim t$ in the present situation)
\begin{equation}
f(x)\sim c\,  D_i\left(\sqrt{2} t^{-1/4} (t-x)\right)
\label{appx7}
\end{equation}
for some constant $c=c(t)$.  Using the known asymptotics of $D_i$
we find
\begin{equation}
f(x)\sim c \left(\sqrt{2} t^{-1/4}(t-x)\right)^i \exp\left\lbrace
-{1\over 2} t^{-1/2} (t-x)^2\right\rbrace
\label{appx8}
\end{equation}
Comparing this with (\ref{appx4}) gives
\begin{equation}
c\sim {2^{-{5\over 2}i-2}\over \sqrt{\pi}} t^{-{3\over 4}i - {3\over 4}}
\exp\left({4\over 3} t^{3/2}\right)\, .
\label{appx9}
\end{equation}
We also find, using (\ref{appx6}), that $b\sim 1/2\sqrt{\pi}$.
\subsubsection{The asymptotics of $\lambda_i$}
If we look at the asymptotics of $f_i(x)$ (the eigenfunction
associated with the eigenvalue $\mu_i$ of $L$, normalized so that
$f_i(0)=1$) given by (\ref{appx1}), (\ref{appx3}), (\ref{appx5}),
and (\ref{appx8}) we see that the main contribution to
$\int_0^\infty f_i(x)^2\,dx $ comes from any region
\[ t^{-1/4}\vert t-x \vert < \eta(t) \]
as long as $\eta(t)\rightarrow\infty$.  For some such region we have,
by (\ref{appx7}) and  (\ref{appx9}),
\[ f_i(x) \sim {2^{-{5\over 2}i-2}\over\sqrt\pi} t^{-{3\over 4}i - {3\over 4}}
D_i\left(\sqrt{2} t^{-1/4}(t-x)\right) \, .\]
Using this, and the fact $\int_{-\infty}^\infty D_i(x)\, dx =\sqrt{2\pi} i!$,
we deduce
\begin{equation}
\int_0^\infty f_i(x)^2\, dx \sim {i!\over\sqrt\pi} 2^{-5i-4}
t^{-{3\over 2}i-{5\over 4}} \exp\left({8\over 3}t^{3/2}\right)\, .
\label{appx10}
\end{equation}
This was the goal of all that went before.
\par
Since $L$ has simple eigenvalues and commutes with our integral operator
$K$ (translated to $(0,\infty)$) the set $\lbrace f_i\rbrace$ is the
set of eigenfunctions of $K$ corresponding to its  eigenvalues
$\lbrace \lambda_i\rbrace$.  However $f_i$ corresponds to $\mu_i$
and there is no assurance that the corresponding eigenvalue of $K$
is its $i^{\rm th}$ largest, which we have denoted by $\lambda_i$.
We shall eventually show that this is actually the case. The first
lemma of this section shows that the eigenvalues of $K$ are simple,
and this is crucial for the argument we shall later use.  (The proof
of the corresponding result for the sine kernel is in \cite{slepian_pollak}.)
\begin{lemma}
If $f_1$ and $f_2$ are eigenfunctions of $L$ corresponding to distinct
eigenvalues, then they are eigenfunctions of $K$ corresponding to
distinct eigenvalues.
\end{lemma}
\par
\noindent Proof.  The eigenvalues of $K$ are the squares of the eigenvalues
of the operator on $(0,\infty)$ with kernel $A(x+y-2t)$.  So
we have to show that if for some $\nu$ we have either
\begin{equation}
\int_0^\infty A(x+y-2t) f_i(y)\, dy = \nu f_i(x) \ \ \ (i=1,2)
\label{eig1}
\end{equation}
or
\begin{equation}
\int_0^\infty A(x+y-2t) f_1(y)=\nu f_1(x)\, , \ \ \int_0^\infty
A(x+y-2t) f_2(y)\, dy = -\nu f_2(x)
\label{eig2}
\end{equation}
then $f_1=f_2$.  (Recall that all eigenfunctions are normalized
to satisfy $f(0)=1$.)
\par
Assuming first that (\ref{eig1}) holds, apply $d^2/dx^2$ to both sides
of the identity with $i=1$ and integrate the resulting integral
by parts twice.  What results is
\[ \nu f_1^{\prime\prime}(x)=-A^\prime(x-2t)+A(x-2t)f_1^\prime(0)+
\int_0^\infty A(x+y-2t) f_1^{\prime\prime}(y)\, dy\, .\]
If we multiply both sides by $f_2(x)$ and integrate we obtain,
using (\ref{eig1}) with $i=2$ (and its differentiated version),
\[ \nu \int_0^\infty f_1^{\prime\prime}(x) f_2(x)\, dx =
-\nu f_2^\prime(0)+\nu f_1^\prime(0) +\nu\int_0^\infty f_2(y)
f_1^{\prime\prime}(y)\, dy\, .\]
Thus, since $\nu\neq 0$ (this follows easily from the fact
that the Fourier transform of $A(x)$ is nonzero), we have
$f_1^\prime(0)=f_2^\prime(0)$.  But it is clear from (\ref{eig_eqn1})
that for any eigenfunction $f$ we have $f^\prime(0)=t^2-\mu$.  Thus
$f_1$ and $f_2$ correspond to the same eigenvalue of $L$ and so
are equal.
\par
The case when (\ref{eig2}) holds is even easier.  Differentiating
both sides of the first relation once and integrating by parts
gives
\[ \nu f_1^\prime(x) = - A(x-2t)-\int_0^\infty A(x+y-2t) f_1^\prime(y)
\, dy\, . \]
Multiplying both sides of this by $f_2(x)$ and integrating, using
the second relation of (\ref{eig2}), we obtain
\[ \nu \int_0^\infty f_1^\prime(x) f_2(x)\, dx =
-\nu+\nu \int_0^\infty f_2(y) f_1^\prime(y)\, dy\, , \]
contradicting $\nu\neq 0$.
\qed
\par
Since, as we now know, the eigenvalues of $K$ are simple, there is
a permutation $\sigma$ of ${\rm{\bf N}}=\lbrace 0,1,2,\cdots\rbrace$
such that $f_i$ is the eigenfunction of $K$ corresponding to
its $\sigma(i)^{th}$ largest eigenvalue $\lambda_{\sigma(i)}$;
this permutation is independent of $t$ or else, by the continuity
of the eigenvalues in $t$, there would be a multiple eigenvalue
for some $t$.
\par
Up to now our kernel $K$ corresponded to an
arbitrary factor $\lambda$.  Of course formula (\ref{eigs_asy}) refers
to the case $\lambda=1$.
\begin{lemma}
\label{lemma:two}
For each $i$ we have $\lambda_i\rightarrow 1$ as $s\rightarrow -\infty$.
\end{lemma}
\par
\noindent Proof. In this case $\lambda=1$ formula (\ref{K_identity}) reads
\begin{equation}
K(x,y)=\int_0^\infty {\rm Ai}(x+z) {\rm Ai}(z+y) \, dz \, .
\label{K_identity2}
\end{equation}
For the purposes of this proof we denote by $A$ (resp.\ $K$) the
operator on $(-\infty,\infty)$ with kernel ${\rm Ai}(x+y)$
(resp.\ $K(x,y)$) and by $P_s$ the projection from $L_2(-\infty,\infty)$
to $L_2(s,\infty)$.  So the $\lambda_i$ are the eigenvalues of
$P_s K P_s$.  Now it is known \cite{hastings_mcleod} that $A^2=I$,
and (\ref{K_identity2}) says that $K=A P_0 A $.  Thus
$K^2=A P_0 A^2 P_0 A = A P_0 A = K$.  In other words, $K$ is a
projection operator.  (This is not surprising since its kernel
is a scaling limit of the kernels  $K_N(x,y)$ of projection operators.)\ \
Clearly it has infinite-dimensional range.  It now follows easily
from the minimax characterization of the eigenvalues that, for
each $i$, the $i^{\rm th}$ largest eigenvalue of $P_s K P_s$
tends to $1$ as $s\rightarrow -\infty$.
\qed
\par
We now state a lemma which is the basis of a trick used by
Fuchs \cite{fuchs64} in his derivation of the asymptotics of the
eigenvalues of the sine kernel.
\begin{lemma}
Let $J(x,y)$ be a symmetric kernel, on a fixed interval, depending
smoothly on a parameter $t$.  Let $g$ be an eigenfunction
normalized so that $\int g(x)^2\, dx = 1$ and let $\lambda$ be
the corresponding eigenvalue.  Then, with the subscript $t$
denoting $\partial/\partial t$, we have
\[ \lambda_t = \int\int J_t(x,y) g(y) g(x) \,dydx\, . \]
\end{lemma}
\par
\noindent Proof.  From
\begin{equation}
\int J(x,y) g(y)\, dy = \lambda g(x)
\label{J_eqn}
\end{equation}
we get
\[ \int J_t(x,y) g(y)\, dy +\int J(x,y) g_t(y)\, dy =
\lambda_t g(x)+  \lambda g_t(x) \, .\]
Multiplying both sides by $g(x)$ and integrating over $x$, we obtain
(using the normalization of $g$ and (\ref{J_eqn}))
\begin{eqnarray*}
\int\int J_t(x,y) g(y) g(x)\, dydx + \int\int J(x,y) g_t(y) g(x) \, dx dy
&=&\lambda_t + \lambda \int g_t(x) g(x)\, dx \\
&=&\lambda_t + \int\int J(x,y) g(y) g_t(x)\, dy dx\, .
\end{eqnarray*}
By symmetry of $J$ the two double integrals involving it cancel,
and we obtain the statement of the lemma.
\qed
\par
We apply the lemma to the opertor with kernel $A(x+y-2t)$
on $(0,\infty)$ whose square is our Airy operator (translated to
act on $(0,\infty)$).  Thus we may write its eigenvalues as
$\sqrt{\lambda_{\sigma(i)}}$, where the square roots might have
either sign.  The lemma gives
\def\ls{\lambda_{\sigma(i)}}
\begin{equation}
{\partial\over\partial t}\sqrt{\ls} = -2 \int_0^\infty\int_0^\infty
A^\prime(x+y-2t) g_i(y) g_i(x)\, dy dx
\label{lambda_eqn}
\end{equation}
where
\[ g_i(x) = f_i(x)\biggl/   \left\lbrace \int_0^\infty f_i(y)^2\, dy
\right\rbrace^{1/2}\, .
\]
But from
\[ \int_0^\infty A(x+y-2t) g_i(y)\, dy = \sqrt{\ls} g_i(x) \]
we obtain upon differentiation that
\[ \int_0^\infty A^\prime(x+y-2t) g_i(y)\, dy = \sqrt{\ls} g_i^\prime(x)\]
so (\ref{lambda_eqn}) may be rewritten
\[
{\partial\over\partial t}\sqrt{\ls} = -2 \sqrt{\ls}
\int_0^\infty g_i^\prime(x) g_i(x)\, dx\, .\]
The integral is of course $-{1\over 2} g_i(0)^2$.  Recalling that
$f_i(0)=1$ we see that we have established the relation
\[
{\partial\over\partial t}\sqrt{\ls} =\sqrt{\ls}\biggl/ \left\lbrace
\int_0^\infty f_i(x)^2\, dx \right\rbrace \, ,
\]
or
\[ {\partial\over\partial t}\log\ls = 2\biggl/ \left\lbrace
\int_0^\infty f_i(x)^2\, dx \right\rbrace \, \]
\par
We now use (\ref{appx10}) and integrate the resulting relation
with respect to $t$ from $t$ to $\infty$.  Recalling
Lemma \ref{lemma:two}, we deduce
\[ -\log\ls \sim {\sqrt\pi\over i!} 2^{5i+3} t^{{3\over 2}i+{3\over 4}}
\exp\left(-{8\over 3} t^{3/2}\right)\, , \]
precisely the right side of (\ref{eigs_asy}).
\par
All that remains now is to show that $\sigma(i)=i$ for all $i$.
But it is clear from this asymptotic relation that $i<j$
implies $\lambda_{\sigma(i)}>\lambda_{\sigma(j)}$ for large $t$
(and so for all $t$), so that $\sigma(i)<\sigma(j)$.  Thus $\sigma$
is order preserving, and since $\sigma:{\rm{\bf  N}}\rightarrow {\rm{\bf  N}}$
is onto we must have $\sigma(i)=i$ for all $i$.
\par
This concludes the derivation of (\ref{eigs_asy}).  The deduction
of (\ref{r_asy}) from (\ref{eigs_asy}) is completely analogous
to the corresponding deduction for  the sine kernel in ref.~\cite{tw1},
Sec.\ VIIB.

\acknowledgments
We wish to thank E.~Br{\'e}zin,   P.~J.~Forrester,
F.~A.~Gr{\"u}nbaum and J.~Harnad  for helpful
comments.  This work was supported in part by the National Science
Foundation, DMS--9001794 and DMS--9216203, and this support is
gratefully acknowledged.

\figure{The probabilities $E(0;s)$ and $E(1;s)$. Of course,
$E(0;s)\rightarrow 1$ as $s\rightarrow +\infty$.}
\figure {The probability densities $F(0;s)$ and $F(1;s)$.  The expected
position of the largest eigenvalue is approximately $-1.7711$ with
variance $0.8132$.  The expected position of the next-largest eigenvalue
is approximately $-3.6754$ with variance $0.5405$.}

\end{document}